%% file: main.tex
\documentclass{vgtc}                          

\ifpdf
  \pdfoutput=1\relax                   
  \usepackage{graphicx}                
  \DeclareGraphicsExtensions{.pdf,.png,.jpg,.jpeg} 
\else
  \usepackage{graphicx}                
  \DeclareGraphicsExtensions{.eps}     
\fi%

\graphicspath{{figures/}{pictures/}{images/}{./}} 

\usepackage{xspace}

\usepackage{microtype}                 
\PassOptionsToPackage{warn}{textcomp}  
\usepackage{textcomp}                  
\usepackage{mathptmx}                  
\usepackage{times}                     
\usepackage{cite}                      
\usepackage{tabu}                      
\usepackage{booktabs}                  
\usepackage{xcolor}
\usepackage{caption}
\usepackage{subcaption}

\usepackage{balance}

\onlineid{0}

\vgtccategory{Research}

\input{macros.tex}

\title{\vis{}: Detecting Similar Patterns in Annotated Literary Text}

\author{Moshe Schorr\thanks{e-mail: moshe@schorr.org}\\ %
        \scriptsize Technion - Israel Institute of Technology %
\and Oren Mishali\thanks{e-mail: omishali@cs.technion.ac.il}\\ %
     \scriptsize Technion - Israel Institute of Technology %
\and Benny Kimelfeld\thanks{e-mail: bennyk@cs.technion.ac.il}\\ %
     \scriptsize Technion - Israel Institute of Technology %
\and Ophir M\"unz-Manor\thanks{e-mail: ophirmm@openu.ac.il}\\ %
     \scriptsize Open University of Israel} %

\abstract{We present a web-based system called \vis{} aiming to assist literary scholars in detecting repetitive patterns in an annotated textual corpus. Pattern detection is made possible using distant reading visualizations that highlight potentially interesting patterns. In addition, the system uses time-series alignment algorithms, and in particular, dynamic time warping (DTW), to detect patterns \emph{automatically}. We present a case-study where an ancient Hebrew poetry corpus was manually annotated with figurative language devices such as metaphors and similes and then loaded into the system. Preliminary results confirm the effectiveness of the system in analyzing the annotated data and in detecting literary patterns and similarities.}

\CCScatlist{
  \CCScatTwelve{Human-centered computing}{Visualization}{Visualization systems and tools}{Visualization toolkits};
  \CCScatTwelve{Theory of computation}{Design and analysis of algorithms}{Data structures design and analysis}{Pattern matching};
}

\begin{document}


\firstsection{Introduction}

\maketitle

Any text features patterns such as repetition of words, structures, or themes. In literary text this phenomenon is even more pronounced; authors use patterns, implicitly or explicitly, as part of their poetic language and the exposure of these patterns and their analysis are the cornerstone of many literary studies. In this paper we present a web-based system \vis{} that provides visualization and analytical tools for annotated (literary) text. The project brings together literary scholars and computer scientists, who share a deep interest in literary hermeneutics, data visualization and pattern recognition. \vis{} aims to support computerized literary studies, and in particular, detection and analysis of \emph{literary patterns}.

Pattern detection within \vis{} is twofold. First, following \emph{distant reading} practices \cite{conf/vissym/JanickeFCS15}, where visualizations play a key role in highlighting potentially interesting patterns, \vis{} provides a variety of visual charts for literary annotated text. The charts make the rich literary information more accessible, thus help the user to identify patterns in a relatively easy manner. In addition, the system uses time-series alignment algorithms, that attempt to detect literary patterns \emph{automatically}. Time-series alignment \cite{FastDTW,journals/corr/abs-1903-09245} is an important question in many different domains such as bioinformatics, computer vision, and speech recognition. The most commonly used method for time-series alignment is dynamic time warping (DTW) \cite{BERNDT94V}, a method that is used here, for the first time, in a literary context. Ultimately the automatic detection juxtaposes the human vis-\`a-vis the computerized findings and hence enables the refinement of both methods.

We have conducted a case study where the system has been used to detect figurative language patterns in ancient Hebrew poetry.\footnote{Piyyut, from Greek poietes, as in poetry in modern English.} The corpora are Hebrew poems from antiquity, starting with the biblical Book of Psalms and ending with the poetic oeuvre of a eight century CE poet from the Galilee. All in all we are dealing with a few hundreds poems that represent a coherent literary tradition, which is ideal for comparative research. The texts were annotated manually by a literary scholar of ancient Hebrew poetry (Ophir M\"unz-Manor) using \catma{}\footnote{http://catma.de} (Computer Aided Textual Markup Analysis), a web annotation tool that was developed in Hamburg University. \catma{} enables the user to annotate texts according to a set of tags that is defined by the scholar. Once the tagging is done the user can execute queries concerning both the text and the tag set. The tag set that was chosen for the case study entails a tagging of figurative language, first and foremost metaphors and similes.

\catma{} provides basic data visualization facilities, and \vis{} was born from the need for more sophisticated and capable visualization and analysis tools.
By and large there are very few literary-based data visualizations tools and \vis{} seeks to fill that gap. \catma{} is very capable in terms of tagging but not much so in the realm of visualization. On the other hand, a tool like Voyant,\footnote{https://voyant-tools.org/} which is very powerful in visualizing text, lacks the ability to visualize tagged text. One can say, then, that \vis{} tries to combine the strength of these two legacy tools.

Our tool imports automatically the tagged corpora from \catma{}, (re)presents it in various graphs and charts, provides comparative graphical tools for grouping the texts, offers aggregative tools for -inter and intra-corpora relations and offers algorithmic pattern and similarity detection.

In the next section we provide an overview of the \vis{} system, and in Section \ref{sec:case_study} we highlight preliminary results from the case study we have made.

\section{\vis{}: System Overview} \label{sec:system} 
\begin{figure}
\begin{subfigure}{.5\textwidth}
  \centering
  \includegraphics[width=.8\linewidth]{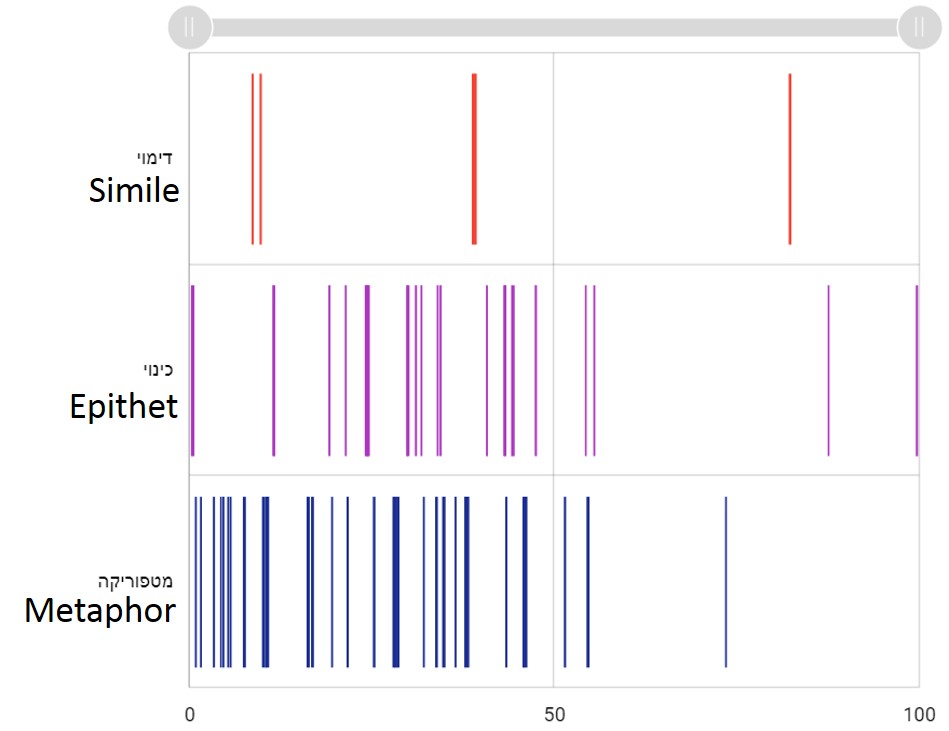}  
  \caption{A Gantt Chart shows density of the tags along the text.}
  \label{fig:gantt}
\end{subfigure}
\begin{subfigure}{.5\textwidth}
  \centering
  \includegraphics[width=.8\linewidth]{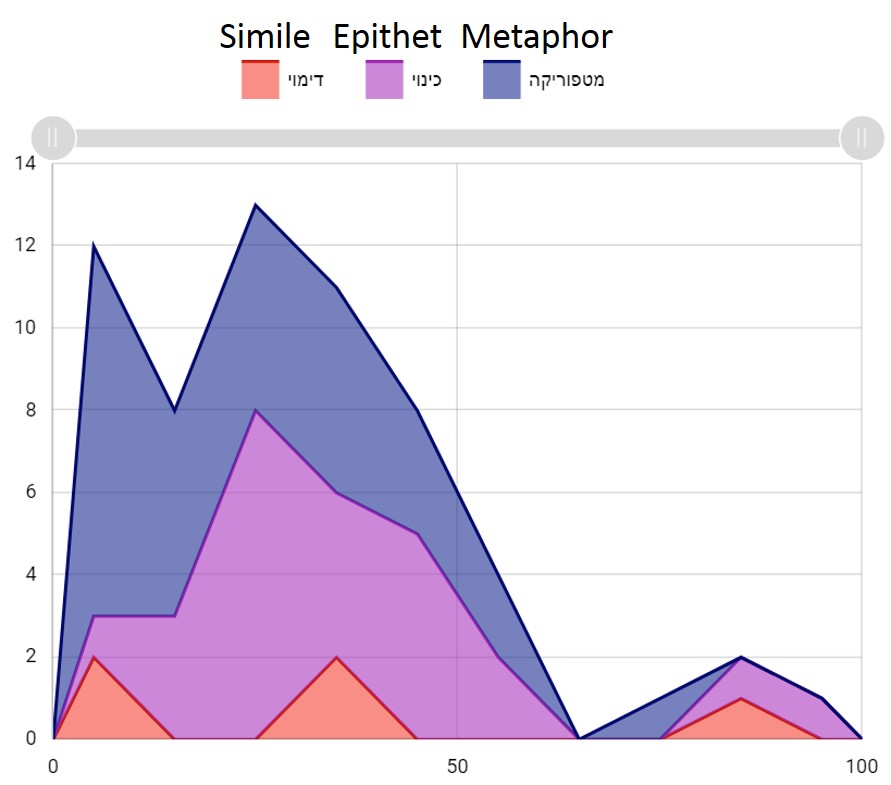}  
  \caption{A Stacked Area Chart shows the tags along the text, emphasizing their quantification.}
  \label{fig:stackedarea}
\end{subfigure}


\begin{subfigure}{.5\textwidth}
  \centering
  \includegraphics[width=.8\linewidth]{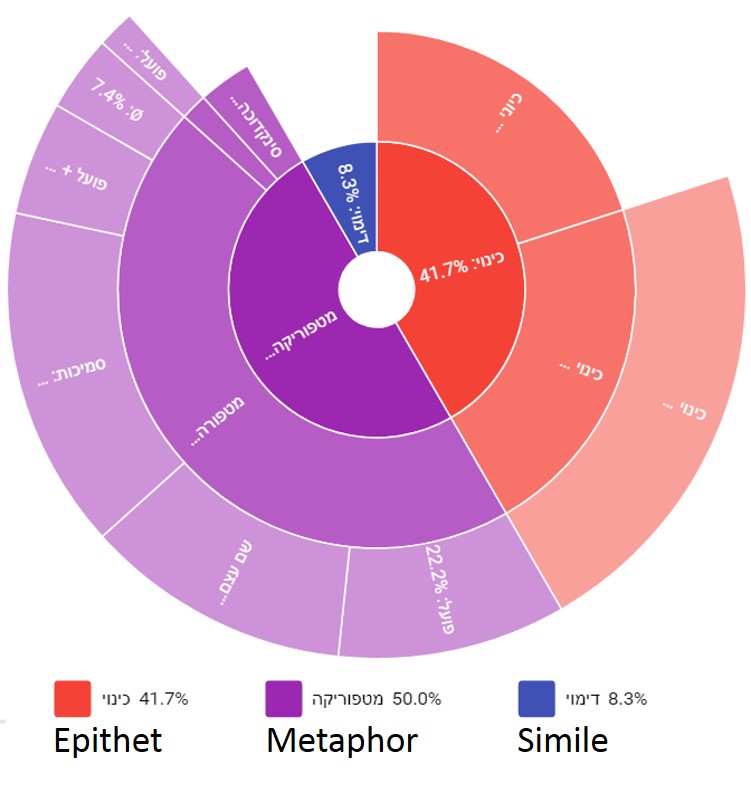}  
  \caption{A Sunburst Chart shows how the literary tags are distributed.}
  \label{fig:sunburst}
\end{subfigure}
\caption{Three visualizations for showing the tags in a specific poem.}
\label{fig:charts_poem}
\end{figure}
The \vis{} system was developed in an agile fashion, emphasizing continuous involvement of the ``customer'' (the literary scholar) in the development process. The development took place in cycles of two weeks, where at the end of each cycle the scholar examined the system, used it on real data, and provided feedback. The feedback helped to refine existing visualizations and to define new ones. For example, during the first iterations, the scholar has found that while the Gantt chart (Figure \ref{fig:gantt}) is useful for showing the density of tags across the text, it does not illustrate the quantitative nature of the data. As a result, the scholar asked the team to insert a new visualization (the Stacked Area Chart in Figure \ref{fig:stackedarea}).

In the rest of this section we explain the architecture of the tool and list key features that help in analyzing similarities between annotated (tagged) texts, in terms of detecting similar patterns.

\subsection{Architecture}
The \vis{} backend is written in Python with Flask and SQLAlchemy. It provides an API for a separate dynamic frontend, which is written in Vue.js. For visualizations we utilize Amcharts v4. Our backend also provides easy connectivity with \catma{} 6, connecting to their GitLab-based backend through user provided API keys. We prompt the user for a key, and then we get information on all user's active projects from \catma{}. We can then allow importing projects or updating projects already imported. The entire process can be done in under a minute from the time the user enters the key until the corpus is ready for viewing. We use a similar data model to \catma{}, although compartmentalized. Users have \emph{Projects}, which are composed of \emph{Texts} and \emph{Annotations}. Annotations associate parts of a text with a \emph{Tag}, and tags belong to a \emph{Tagset}.

\subsection{Distant Reading Features}
\vis{} offers several data visualizations.
On an individual level, we allow seeing the distribution of tags in a Gantt-style chart, showing the locations of tags within the text. The Gantt chart shows the spreading and density of tags in the text, and thus help the user to identify patterns mostly related to the \emph{location} of tags. The user however may also be interested in revealing patterns related to the \emph{quantity} of tags in the text. For that we offer an aggregated Stacked Area chart, as well as a Sunburst chart, showing the proportions and quantities associated with each tag in the text. These three types of charts allow the user to get an immediate visual overview of the tagged features throughout the text, including breakdown, location, and density. They are illustrated in Figure \ref{fig:charts_poem} for an Hebrew poem called \emph{Tehila L'aRam} (translated to: Glory to the Supreme).

The horizontal axis of the Gantt chart (Figure \ref{fig:gantt}), and of the Stacked Area chart (Figure \ref{fig:stackedarea}) shows the location of the tags within the text, where their vertical axis shows the tag type (red for a simile, purple for an epithet, and blue for a metaphor). Both charts reveal that the first half of the poem is rich with epithets and metaphors, while more or less in the middle of the poem there is a sudden drop. We further discuss this observation in Section \ref{sec:case_study}. The third Sunburst chart (Figure \ref{fig:sunburst}) quantifies the tags based on their defined hierarchy. We can see that in this specific poem 50\% of the tags are of type metaphor (purple), 41.7\% are epithets (red), and 8.3\% are similes (blue). The different color shadows are subtypes such as synecdoche, nouns, and verbs.

For our key view, the Gantt, we connect the text and the visualization. That is to say, as the user hovers a tag in the chart, a text pane that is located next to it reacts accordingly and the text scrolls to and highlights that tag. In addition, by changing the bounds of the visualization (using the upper scroll bar shown in Figure \ref{fig:gantt}), the text adjusts to fit those bounds.

\begin{figure}
\centering
\begin{minipage}[t]{\columnwidth}
	\includegraphics[width=\textwidth]{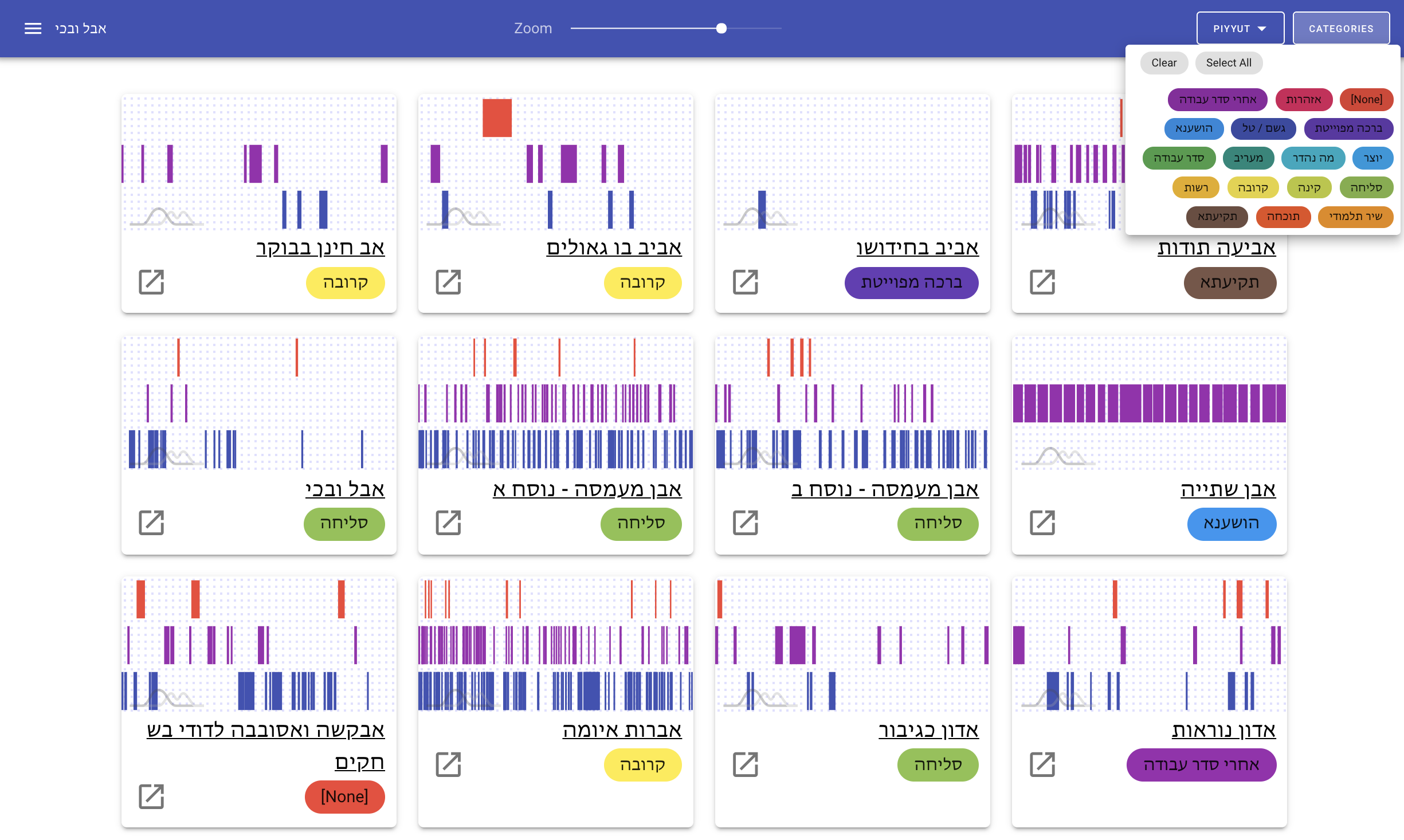}
	\end{minipage}
	\caption{A portion of a gallery view for the pre-classical Piyyut corpus.}
	\label{fig:gallery_view}
\end{figure}

The three charts mentioned above allow to explore the tags in individual texts. Yet for the user to be able to identify repeating patterns across several texts, it should be possible to view the corpus as a whole. This is supported by a \emph{gallery view}, where minimized Gantt charts are laid side by side. A portion of a gallery view for the pre-classical Piyyut corpus is shown in Figure \ref{fig:gallery_view}. The figure shows twelve poems from the corpus, and the user may zoom-out to reveal more charts onto the same page.

Naturally, the next step would be to place texts having similar patterns in corresponding groups.
This is further complemented by our \emph{board} feature, which provides a view similar to that shown in Figure \ref{fig:gallery_view}, but with added interactivity allowing to group texts together, for instance, to group texts that use metaphors heavily at the beginning and end, but not in the middle. It is possible to create named categories and then to place poems in categories by dragging them around. The broader goal here is to provide insight into textual features with simple human-based visual pattern recognition.

\section{Automatic Similarity Detection}
One of the areas we are researching is automatic detection of similarities. While much work has been done on text reuse and other fully computerized methods of evaluating textual similarity, including in Hebrew \cite{ShmidmanKP16}, there has been less work done applying computerized methods to hand-tagged data. The hope is that computerized methods can efficiently leverage the time and effort put into creating the data, and that it can add value beyond what can be done with the plain text.

We apply a novel combination of existing methods to this task. Our initial phase is to vectorize each text for each tag we are using to evaluate the similarity. We convert the tag data to a binary vector, where each element in the vector represents whether the given character in that text has the given tag. Since we do not expect to see exact matches, and we expect to see shifting and warping, as well as comparing across texts of very disparate widths, we need a fuzzier metric. We apply dynamic time warping (DTW) \cite{sakoechiba1978} to the task, an algorithm originally applied to speech signal analysis, however it has since seen a wide range of applications \cite{journals/corr/abs-1903-09245}.

The time parameterization is equally applicable to any similar feature vector. However, when we need to calculate similarities for a substantial number of texts, we then need to calculate similarities for each pair, and the \emph{additional} quadratic calculation of DTW itself becomes prohibitively slow. We therefore use FastDTW\cite{FastDTW} for the underlying alignment, as it provides similar accuracy in linear time. Finally, having aligned our sequences we then score our alignments. We use TAM \cite{TAMFolgado} for a baseline score, as we find it provides substantially better results than simpler methods like Euclidean distance.

Having calculated the TAM for the given tag between a pair of texts, we then change it to a 0 to 1 scale, and apply one last modification of our own, namely, we multiply by a weight factor. We calculate our weight factor as \( min(\frac{10*H(v1)}{|v1|}, 1)*min(\frac{10*H(v2)}{|v2|}, 1) \), applying a linear slope penalty for sparseness of the vectors (where \emph{H} denotes the Hamming weight). We do this since we find that the tags tend to be fairly sparse, and while from a mathematical point of view treating two empty vectors as ``perfectly similar'' is more conventional than our treatment as perfectly dissimilar, given the underlying sparseness this would cause a large amount of useless results (the penalty can be easily neutralized for non-sparse corpora). This also allows a substantial speedup by being able to avoid some calculations.

\begin{figure}[t]
\centering
\begin{minipage}[t]{0.9\columnwidth}
	\includegraphics[width=\textwidth]{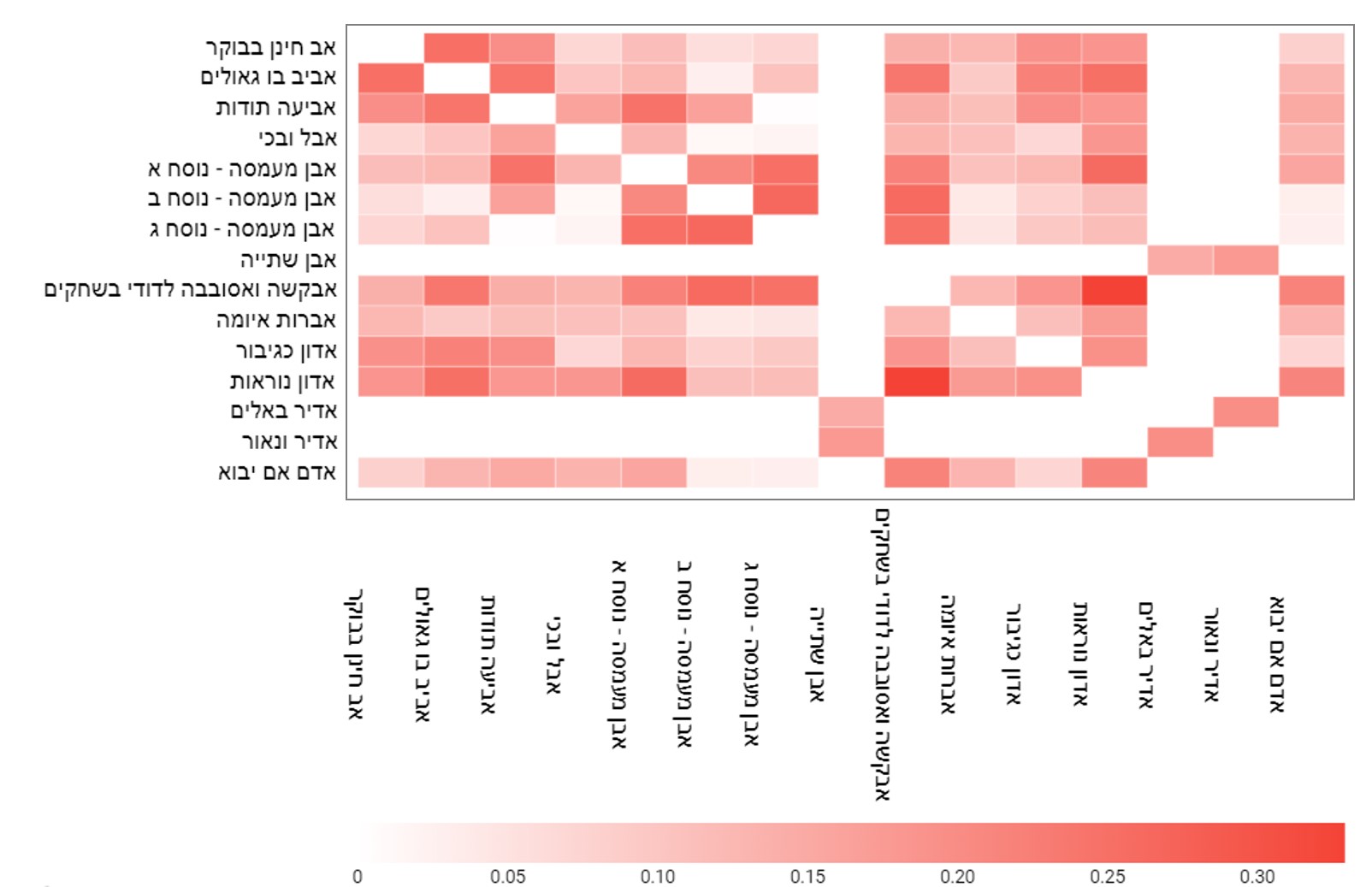}
	\end{minipage}
	\caption{A heatmap showing the similarity between pairs of tagged texts as derived from the DTW alorithm. Both x-axis and y-axis show (the same) Hebrew poems from the pre-classical Piyyut corpus. Each cell vislualizes the level of similarity of a pair. A redder cell indicates a pair of texts with higher tag similarity.}
	\label{fig:heatmap}
\end{figure}

Having completed calculating pairwise for a group of texts we can then derive a heatmap (Figure \ref{fig:heatmap}). We use this visualization to show the relationship between pairs, as well as allowing a view of groupings, emerging from blocks of similar texts.

\section{Case Study}
\label{sec:case_study}
In order to examine \vis{} as a tool for the literary scholar we conducted a short experiment. One section of the corpora under discussion was analyzed manually fifteen years ago by M\"unz-Manor, as part of his doctoral dissertation \cite{ophir2007}. The research focused on the nature of figurative language in the corpus, namely the use of metaphors and similes. The main conclusion of the dissertation was that the corpus under discussion reveals poetic “disinterest” in using figurative language, a most uncommon practice in the universe of Hebrew verse or verse in general \cite{ophir2011}.

\begin{figure}[t]
\centering
\begin{minipage}[t]{0.9\columnwidth}
	\includegraphics[width=\textwidth]{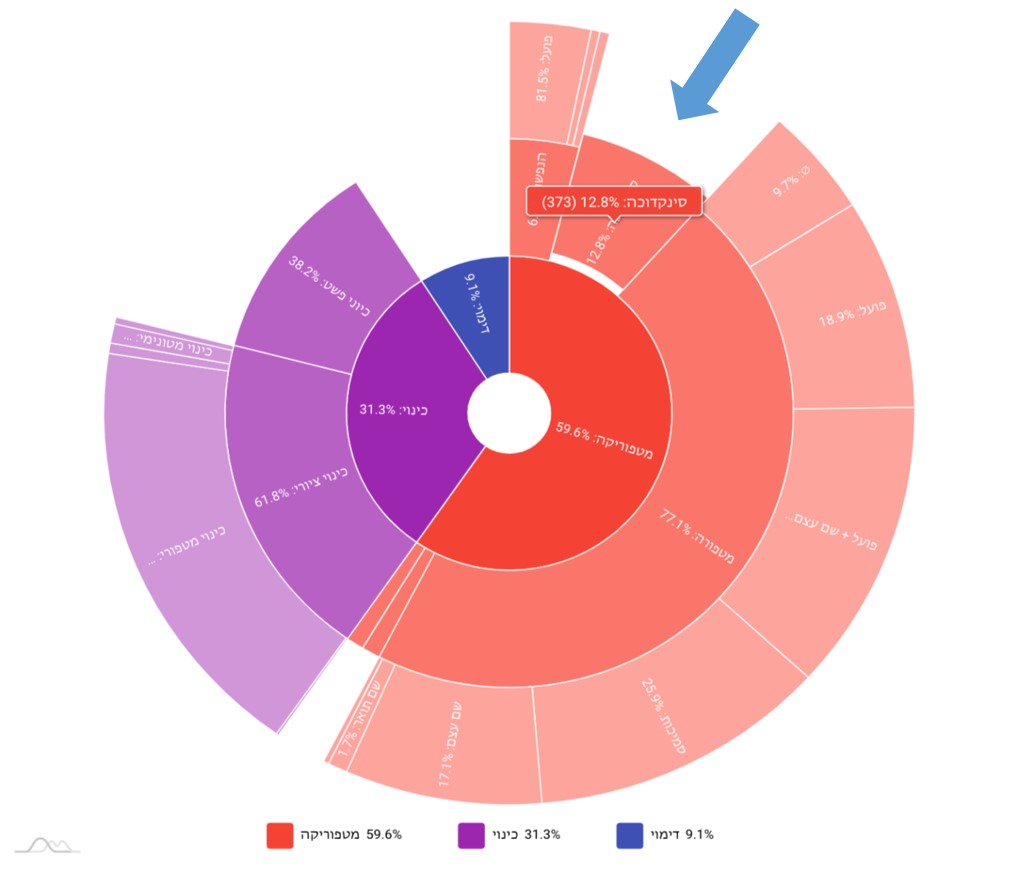}
	\end{minipage}
	\caption{A Sunburst chart showing the distribution of the various figurative devices in the pre-classical Piyyut corpus.}
	\label{fig:sunburst_corpus}
\end{figure}

We wanted to come back to the results of the dissertation, this time using our tool, in order to corroborate or refute the original findings. In order to do so we transferred the original manual taggings from the dissertation into \catma{} and then exported them to \vis{}. By that we were able to use data visualization in order to have a fresh look at the corpus. Some of the preliminary results were revealing; for example, the Sunburst chart (Figure \ref{fig:sunburst_corpus}) helped us refine the statistical distribution of the various figurative devices. On the one hand we were able to corroborate the finding that the simile is rather marginal in the corpus \cite{ophir2009}, a point that was already noted in the doctoral dissertation, and on the other that the synecdoche is much more common than was initially thought (marked with an arrow in the figure). In another case we were able to detect a specific poetic genre (the Hosanna poems) in which figurative language was much more pronounced than expected.

\begin{figure}[t]
\centering
\begin{minipage}[t]{0.9\columnwidth}
	\includegraphics[width=\textwidth]{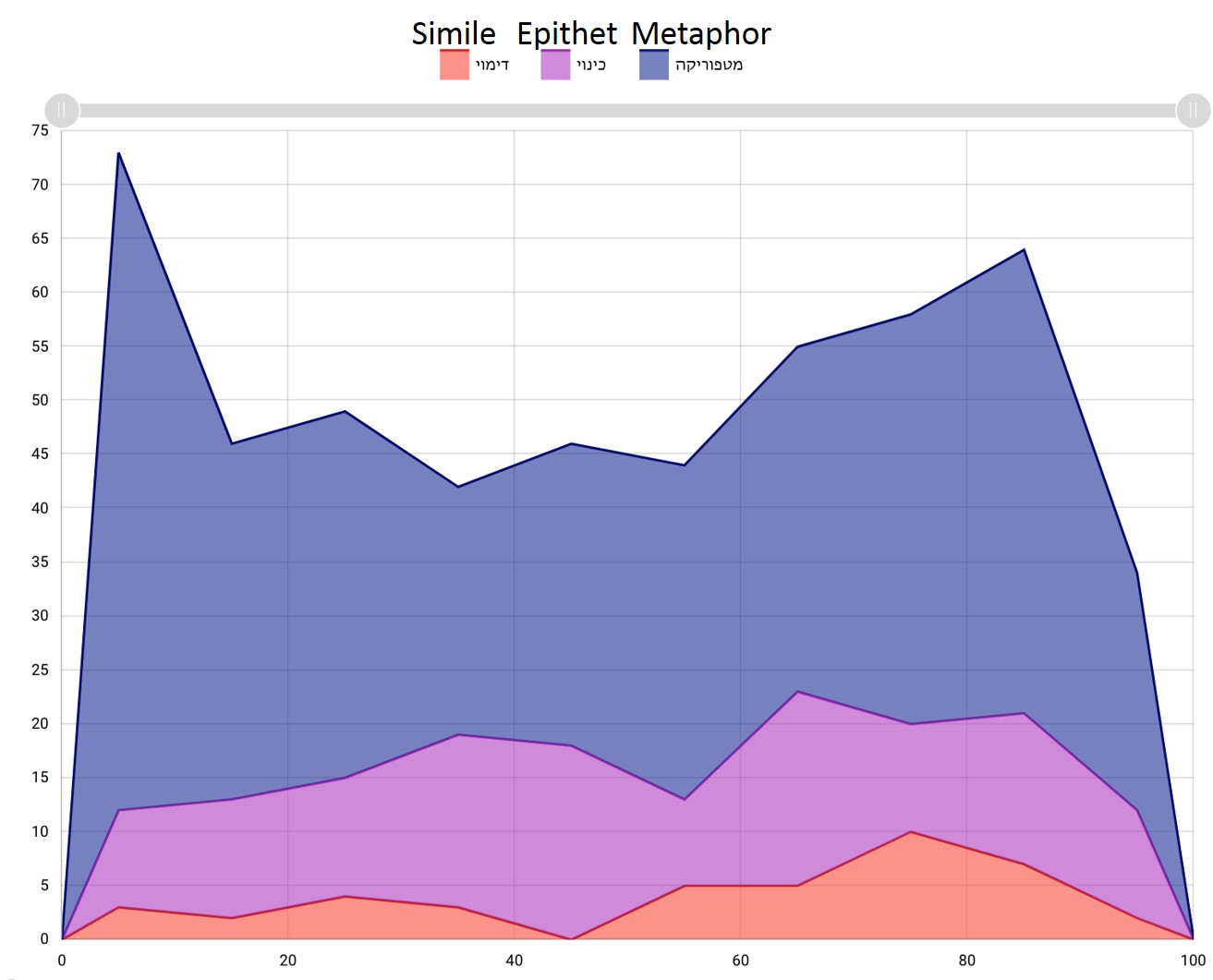}
	\end{minipage}
	\caption{A poem of the Avodah genre where figurative devices appear consistently along the text. This exceptional pattern helped to identify the poem as belonging to a later period.}
	\label{fig:avoda_poem}
\end{figure}

The \vis{} tool also helped to detect patterns that went unnoticed in traditional literary studies of the corpora, and to single out an important poetical phenomenon. It turns out that in one specific genre, the Avodah poems for the Day of Atonement \cite{swartz2005}, there is a sudden drop in the usage of metaphors and other figurative devices at a certain point in the poem followed by a small resurgence of figurative language towards the end of the poem (as illustrated in Figure \ref{fig:gantt} and \ref{fig:stackedarea} for a specific poem in the genre). Once we detected the fact that this pattern is consistent only in this group of poems we went back to the texts themselves and it was revealed that a change in the narrative governs this drop. Interestingly, deviation from this pattern enabled us to single out a poem of the Avodah genre which is much later from the rest of the poems (Figure \ref{fig:avoda_poem}). That is to say, a change in the pattern detected signified a poetic change in the text.

We took then another step and annotated, again by traditional human hermeneutical tagging, the Book of Psalms. Once this corpus of ancient Hebrew poetry was imported to our tool we were able to compare the findings concerning figurative language in both corpora. Figurative language is very common in the biblical book and its juxtaposition with the later corpus revealed as well meaningful differences that we would like to examine in the near future. 

\begin{figure}[t]
\centering
\begin{minipage}[t]{0.9\columnwidth}
	\includegraphics[width=\textwidth]{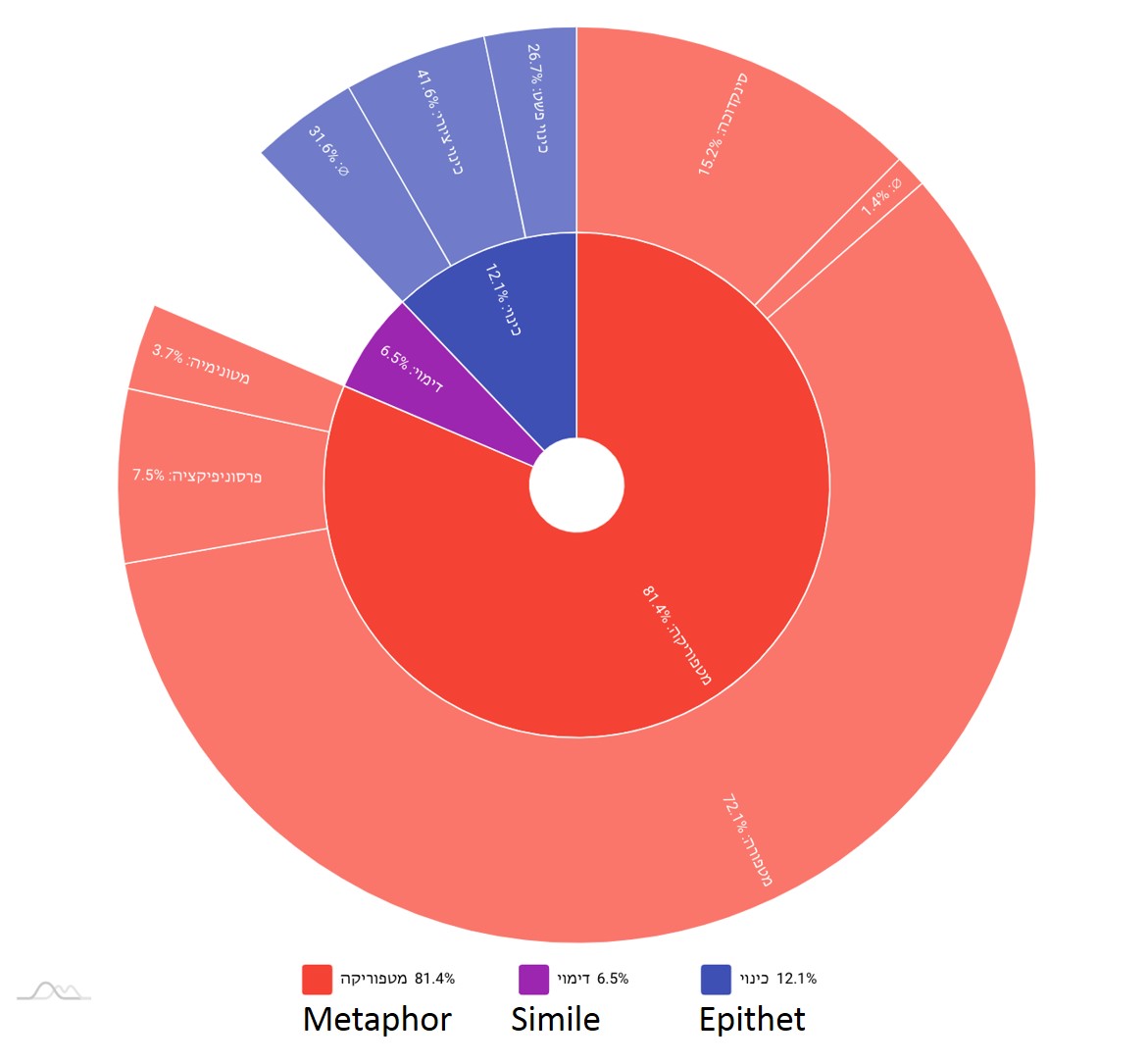}
	\end{minipage}
	\caption{Distribution of figurative devices in the Book of Psalms. Comparing these findings to the pre-classical Piyyut corpus in Figure \ref{fig:sunburst_corpus} revealed meaningful differences. For example, in the Psalms 81\% of the figurative devices are metaphors whereas in Piyyut the respective figure is  60\%. Namely, the density of metaphors in the two corpora differs considerably, a fact that goes hand in hand with other poetic characteristics of the corpora.}
	\label{fig:sunburst_tehilim}
\end{figure}

In order to evaluate the quality of the similarity algorithm we have added a user rating mechanism to the \vis{} tool. The mechanism is implemented using our board feature, that is described in Section \ref{sec:system}. We created a list of test groups based on our automatic scores. Each group consisted of a \emph{target} document, and five other candidate documents that were created with the three most similar documents to the target, one moderately similar (chosen from 4th-10th place), and one drawn at random. The literary scholar (Ophir M\"unz-Manor) was then presented with each group and was asked to rank the five documents in order of similarity to the target document. In order to avoid priming, the five documents were presented randomly.

The results of our initial evaluation were promising. While the algorithm was unable to precisely rank them, it correctly ascertained the groupings in most cases. Most promisingly, the least similar document was correctly identified roughly 70\% of the time (vs. 20\% by random chance). Given the specifics of the method where the least similar by the metrics was substantively different, being randomly chosen and not chosen for being of interest, it suggests that it was indeed capable of differentiating between similar and dissimilar texts. We also note that the scholar himself found the nuances of the choices to be difficult and unclear at times. In short, we believe that the system as it currently exists is already capable of suggesting preliminary matches and groupings of interest.

\section{Conclusion}
All in all, the innovative nature of the group of scholars and the project relates to the combination of “traditional” literary scholarship with state-of-the-art visualization and algorithmic tools and procedures. Moreover, the project facilitates discussions concerning the opportunities and limitations of literary data in a big-data setting, the meaning of the so-called “distant reading” in this context, the interaction between large-scale visualization and particular textual occurrences, and, finally, the similarities and differences between human and computerized “reading”.

Another unique feature of the project lies in the fact that we visualize and detect patterns not in the texts themselves but rather on the level of tagging. Since our tags are based at the moment on manual annotation of the literary texts we have access to the “hermeneutical stratum”, namely to a “second-order” data of human provenance. In other words, patterns exist not only in the text but also in the attached interpretation thereof.

For us, both the primary text and the tagged text are data and as such both are interesting. While the former can be called first-order data the latter might be called second-order data. That said, from a humanistic perspective the text and its interpretation(s) are both sides of one coin and we have interest in both of them. For that reason in the next stage we plan to add visualization capabilities of both the text and the tags, hence improving the pattern recognition abilities of the tool. One of the advantages of \vis{} will be its ability to visualize and analyze the correspondence between the two strata. Another feature would enable the juxtaposition of several human interpretations on the same text hence enabling also a comparative study of human hermeneutical multiplicity.

We have currently tested \vis{} on literary text with figure of speech tags. As a future work, we plan to use the system with a greater variety of literary tags, as well as to explore its applicability for more domains.

\balance 

\bibliographystyle{abbrv}

\bibliography{template}
\end{document}

%% file: macros.tex
\newcommand*{\vis}{\textsl{ViS-\`A-ViS}\xspace}
\newcommand*{\catma}{\textsc{CATMA}\xspace}

%% file: main.bbl
\begin{thebibliography}{10}

\bibitem{BERNDT94V}
D.~J. Berndt and J.~Clifford.
\newblock Using dynamic time warping to find patterns in time series.
\newblock In {\em AAAI Workshop on Knowledge Discovery in Databases}, pages
  359--370, Seattle, Washington, July 1994.

\bibitem{TAMFolgado}
D.~Folgado, M.~Barandas, R.~Matias, R.~Martins, M.~Carvalho, and H.~Gamboa.
\newblock Time alignment measurement for time series.
\newblock {\em Pattern Recognition}, 81:268 -- 279, 2018.

\bibitem{conf/vissym/JanickeFCS15}
S.~J{\"a}nicke, G.~Franzini, M.~F. Cheema, and G.~Scheuermann.
\newblock On close and distant reading in digital humanities: A survey and
  future challenges.
\newblock In {\em 17th Eurographics Conference on Visualization, EuroVis},
  pages 83--103. Eurographics Association, 2015.

\bibitem{journals/corr/abs-1903-09245}
S.~Khorram, M.~G. McInnis, and E.~M. Provost.
\newblock Trainable time warping: Aligning time-series in the continuous-time
  domain.
\newblock {\em CoRR}, abs/1903.09245, 2019.

\bibitem{ophir2007}
O.~M\"unz-Manor.
\newblock {\em Entitled Studies in Figurative Language of Pre-Classical
  Piyyut}.
\newblock PhD thesis, The Hebrew University of Jerusaelm, 2007 (Hebrew).

\bibitem{ophir2009}
O.~M\"unz-Manor.
\newblock As the apple among fruits, so the priest when he emerges: Poetic
  similies in pre-classical poems of the 'how lovely' genre.
\newblock In {\em Ginzei Qedem - Genizah Research Annual 5}, pages 165--188,
  2009.

\bibitem{ophir2011}
O.~M\"unz-Manor.
\newblock Figurative language in early piyyut.
\newblock In {\em Giving a Diamond: Essays in Honor of Joseph Yahalom on the
  Occasion of His Seventieth Birthday}, pages 51--68, 2011.

\bibitem{sakoechiba1978}
H.~Sakoe and S.~Chiba.
\newblock Dynamic programming algorithm optimization for spoken word
  recognition.
\newblock {\em IEEE transactions on acoustics, speech, and signal processing},
  26(1):43--49, 1978.

\bibitem{FastDTW}
S.~Salvador and P.~C. 0001.
\newblock Toward accurate dynamic time warping in linear time and space.
\newblock {\em Intell. Data Anal}, 11(5):561--580, 2007.

\bibitem{ShmidmanKP16}
A.~Shmidman, M.~Koppel, and E.~Porat.
\newblock Identification of parallel passages across a large hebrew/aramaic
  corpus.
\newblock {\em CoRR}, abs/1602.08715, 2016.

\bibitem{swartz2005}
M.~Swartz and J.~Yahalom.
\newblock Avodah: An anthology of ancient poetry for yom kippur.
\newblock In {\em University Park: The Pennsylvania State University Press},
  2005.

\end{thebibliography}
